\documentclass[aps,prl,letterpaper,twocolumn,10pt,superscriptaddress]{revtex4-1}

\usepackage[utf8]{inputenc}	

\usepackage{hyperref} 
\hypersetup{ 
     hyperindex=true, 		
     colorlinks=true, 		
     urlcolor= Blue3,  		
     linkcolor= Blue3, 		
     citecolor=Green4,		
}

\usepackage{amssymb} 		
\usepackage{amsfonts}		
\usepackage{amsmath}        
\usepackage{bm}				
\usepackage{mathrsfs}		
\usepackage{graphicx}		
\usepackage[x11names,svgnames]{xcolor}       
\usepackage{units}					
\usepackage{color}		

\begin{document}
\title{Pauli-Heisenberg Blockade of Electron Quantum Transport}
\author{Karl Thibault}
\email{Karl.Thibault@USherbrooke.ca}
\affiliation{Département de physique, Université de Sherbrooke, Sherbrooke, Québec, J1K 2R1, Canada}
\author{Julien Gabelli}
\email{Julien.Gabelli@U-PSud.fr}
\affiliation{Laboratoire de Physique des Solides, Univ. Paris-Sud, CNRS, UMR 8502, F-91405 Orsay Cedex, France}
\author{Christian Lupien}
\email{Christian.Lupien@USherbrooke.ca}
\affiliation{Département de physique, Université de Sherbrooke, Sherbrooke, Québec, J1K 2R1, Canada}
\author{Bertrand Reulet}
\email{Bertrand.Reulet@USherbrooke.ca}
\affiliation{Département de physique, Université de Sherbrooke, Sherbrooke, Québec, J1K 2R1, Canada}
\date{\today}
\maketitle

\textbf{Conduction of electrons in matter is ultimately described by quantum mechanics. Yet at low frequency or long time scales, low temperature quantum transport is perfectly described by this very simple idea: electrons are emitted by the contacts into the sample which they may cross with a finite probability \cite{Martin1992,Buttiker1992}. Combined with Fermi statistics, this partition of the electron flow accounts for the full statistics of electron transport \cite{Lesovik1994}. When it comes to short time scales, a key question must be clarified: are there correlations between successive attempts of the electrons to cross the sample? While there are theoretical predictions \cite{Martin1992} and several experimental indications for the existence of such correlations \cite{Reznikov1995,Kumar1996}, no direct experimental evidence has ever been provided. Here we show a direct experimental proof of how temperature and voltage bias control the electron flow: while temperature $T$ leads to a jitter which tends to decorrelate electron transport after a time $\hbar/k_BT$, the bias voltage $V$ induces strong correlations/anticorrelations which oscillate with a period $h/eV$. Our experiment reveals how time scales related to voltage and temperature operate on quantum transport in a coherent conductor. In complex quantum systems, the method we have developed might offer direct access to other relevant time scales related, for example, to internal dynamics, coupling to other degrees of freedom, or correlations between electrons.}
\begin{figure}[t!]\centering
\includegraphics[width=0.9\columnwidth]{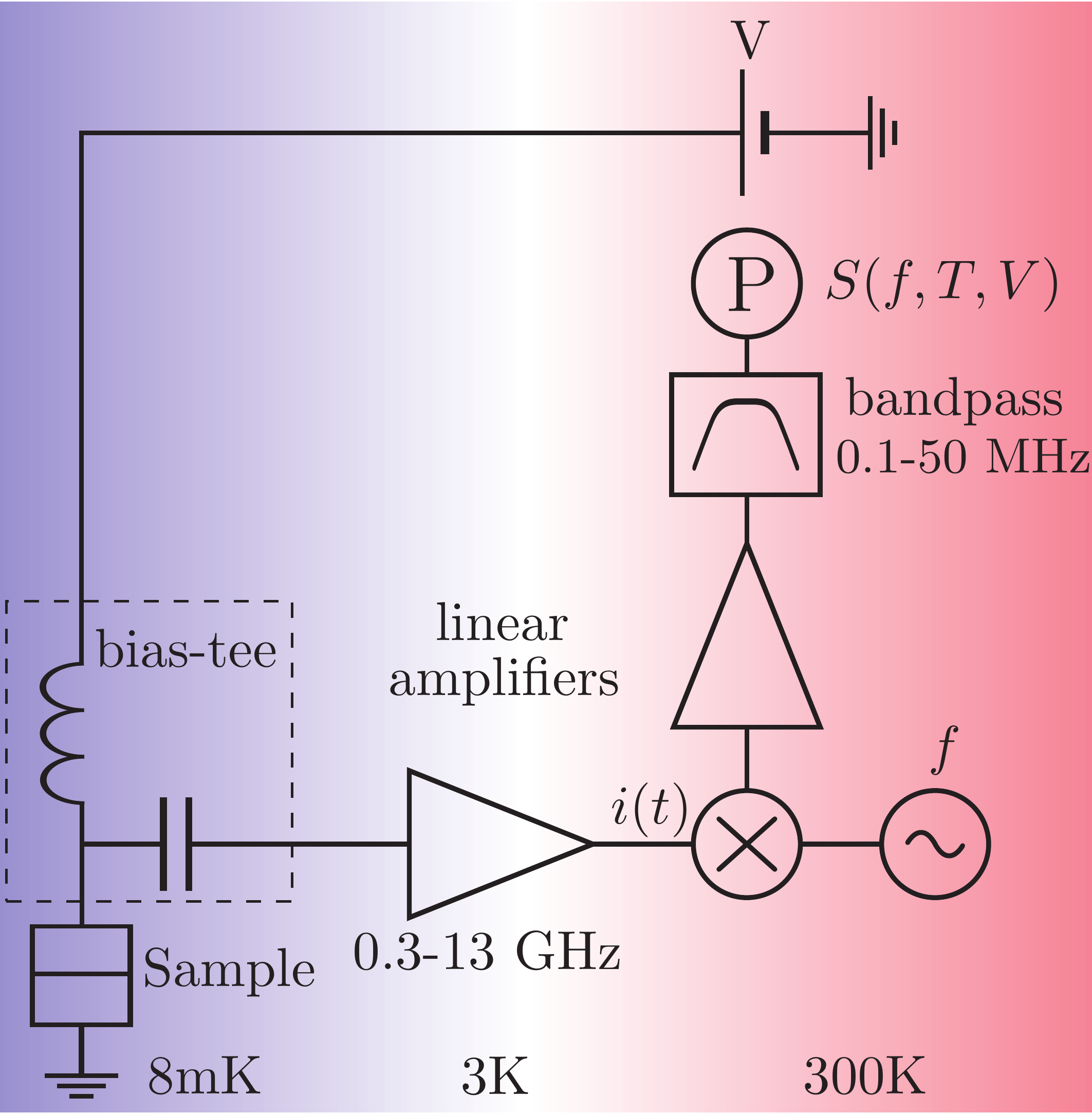}
\caption[]{\textbf{Experimental setup.} An Al/Al oxide/Al tunnel junction of resistance $R=51$ $\Omega$ is cooled to 8 mK in a dilution refrigerator. A perpendicular magnetic field is applied to keep the Al in its normal state. A resistive heater on the cold plate allows us to vary the temperature of the sample. The sample is dc biased through the dc port of a bias-tee. The current fluctuations generated by the junction are amplified in the frequency range 0.3-13 GHz by a cryogenic high-electron-mobility transistor amplifier placed at 3 K. The resulting signal is down converted, using a frequency mixer (symbol \textcircled{\tiny{\textsf{X}}}), to low frequency by multiplication with a local oscillator of variable frequency $f$, then band-pass filtered between 0.1-50 Mhz. Using a power detector (symbol \textcircled{\tiny{P}}), we measure the power of that signal which is given by $P(f)= G_A(f)[S(f)+S_A(f)]\Delta f$, where $G_A(f)$ is the total gain of the measurement system, $S_A(f)$ the noise added by the detection and $S(f)$ the noise emitted by the sample. $G_A(f)$ contains the effects of amplification, cable attenuation and reflection as well as the frequency dependent coupling between the sample and the microwave circuit. The bandwidth $\Delta f \simeq 100$ MHz is much smaller than all relevant frequency scales so that $G_A(f)$, $S(f)$ and $S_A(f)$ hardly vary within $\Delta f$.}
\label{fig:setup}
\end{figure}

\vspace{\baselineskip}
In order to probe temporal correlations between electrons, we have studied the correlator between current fluctuations $i(t)$ measured at two times separated by $\tau$, $C(\tau)=\langle i(t)i(t+\tau) \rangle$, where $\langle . \rangle$ denotes statistical averaging. We calculate this correlator by Fourier transform of the detected frequency-dependent power spectrum of current fluctuations generated by a tunnel junction placed at very low temperature. The very short time resolution required to access time scales relevant to electron transport is achieved thanks to the ultra-wide bandwidth, 0.3-13 GHz, of our detection setup shown in Fig. \ref{fig:setup}. The calibration procedure can be found in the Methods section.

\vspace{\baselineskip}
\emph{Results.}
The electron temperature in the sample is measured using the shot noise thermometer technique \cite{Spietz2003}, which consists of measuring the zero frequency spectral density of current fluctuations vs. voltage at low frequency, \textit{i.e.} $hf \ll k_BT$, and fitting it with the known classical formula \cite{Dahm1969} :
\begin{equation}
S(f=0,V,T)=GeV \coth \left( \frac{eV}{2k_BT} \right).
\end{equation}
We obtain an electron temperature $T=35$ mK when the phonon temperature is $T_{ph} = 8$ mK (measured by a thermometer on the cold plate of the refrigerator). For phonons above 50mK, we observe $T = T_{ph}$. We believe the discrepancy between $T$ and $T_{ph}$ at the lowest temperature is due to the emission of noise with very wide bandwidth by the amplifier towards the sample. In the following, the spectral density of current fluctuations is expressed in terms of noise temperature using $T_N(f) = S(f)/(2k_BG)$. 
\begin{figure}[t]\centering
\includegraphics[width=1\columnwidth]{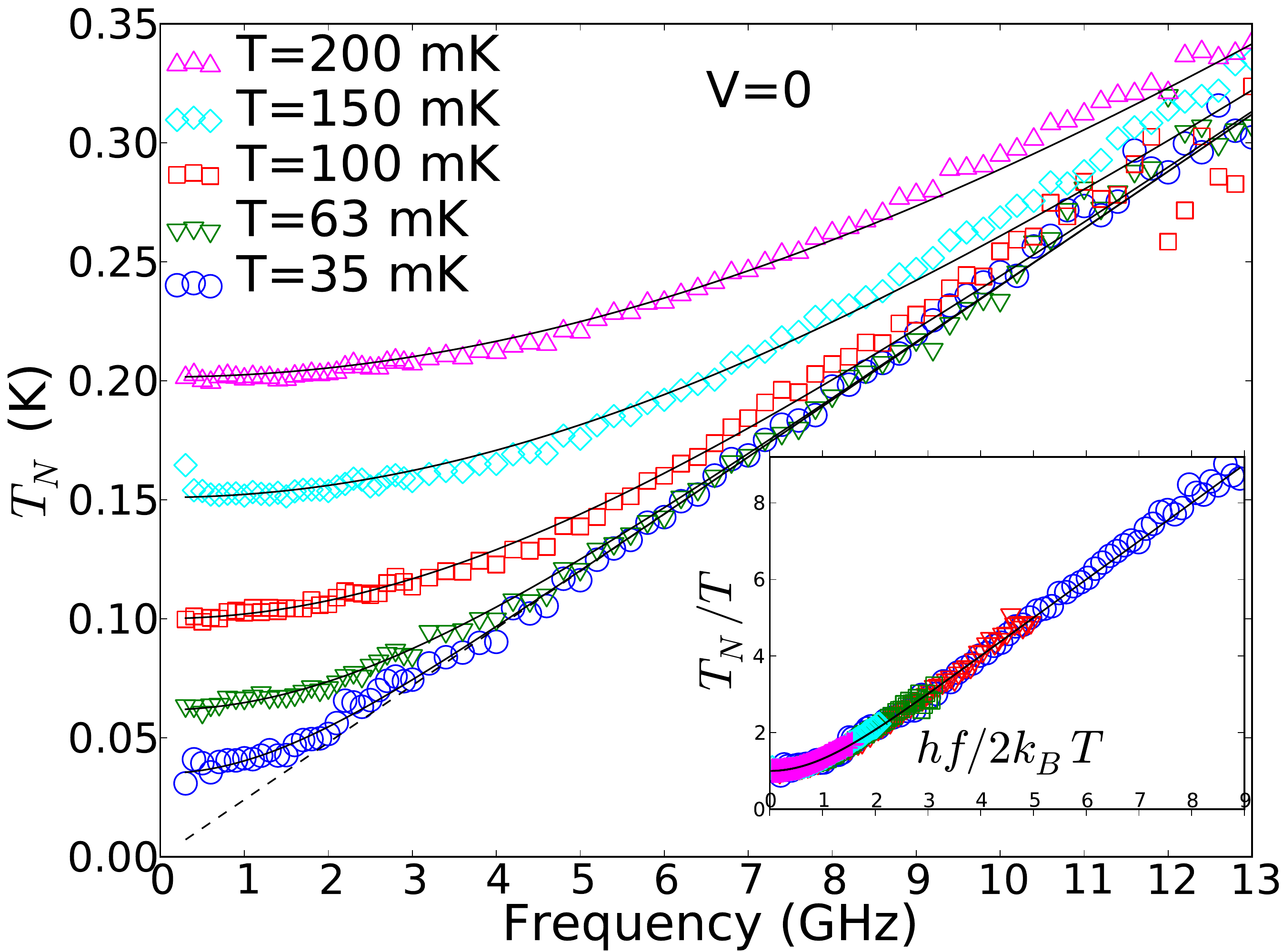}
\caption[]{\textbf{Equilibrium noise temperature vs. frequency for various electron temperatures $T$.} Symbols are experimental data and solid lines are theoretical expectations of equation (\ref{eq:S_eq}). \\ \textbf{Inset} : Experimental rescaled noise temperature $T_N/T$ vs. rescaled frequency $hf/(2k_BT)$.}
\label{fig:S_T}
\end{figure}

\emph{Thermal noise spectroscopy.}
On Fig. \ref{fig:S_T}, we show measurements of $T_N$ vs. frequency for various electron temperatures $T$ between 35 and 200 mK, when the sample is at equilibrium, \textit{i.e.} with no bias ($V=0$). We observe that at low frequency one has $T_N(0) = T$ which is the classical Johnson-Nyquist noise \cite{Johnson1928,Nyquist1928}. At high frequency $hf \gg k_BT$, all curves approach the zero temperature curve (dotted black line) which corresponds to the so-called vacuum fluctuations $S_{vac}(f) = Ghf$. These quantum zero-point fluctuations had previously been characterized as a function of frequency for a resistor \cite{Koch1981, Mariantoni2010} and a superconducting resonator\cite{Basset2010}. The equilibrium spectral density of noise is predicted to be \cite{Callen1951} :
\begin{equation}
S_{eq}(f, T) = Ghf \coth \left( \frac{hf}{2 k_B T} \right). 
\label{eq:S_eq}
\end{equation}
The black lines on Fig. \ref{fig:S_T} represent equation (\ref{eq:S_eq}) with no adjustable parameters. Our data are in very good agreement with the theoretical predictions. According to equation (\ref{eq:S_eq}), the rescaled noise temperature $T_N/T$ is a function of frequency and temperature only via the ratio $hf/(2k_BT)$. We show in the inset of Fig. 2 the measured $T_N/T$ vs. rescaled frequency $hf/(2k_BT)$ for all our data. We indeed observe that all the data collapse on a single curve for a wide interval of $hf/(2k_BT)$ between 0.075 and 9.
\begin{figure}[t]\centering
\includegraphics[width=1\columnwidth]{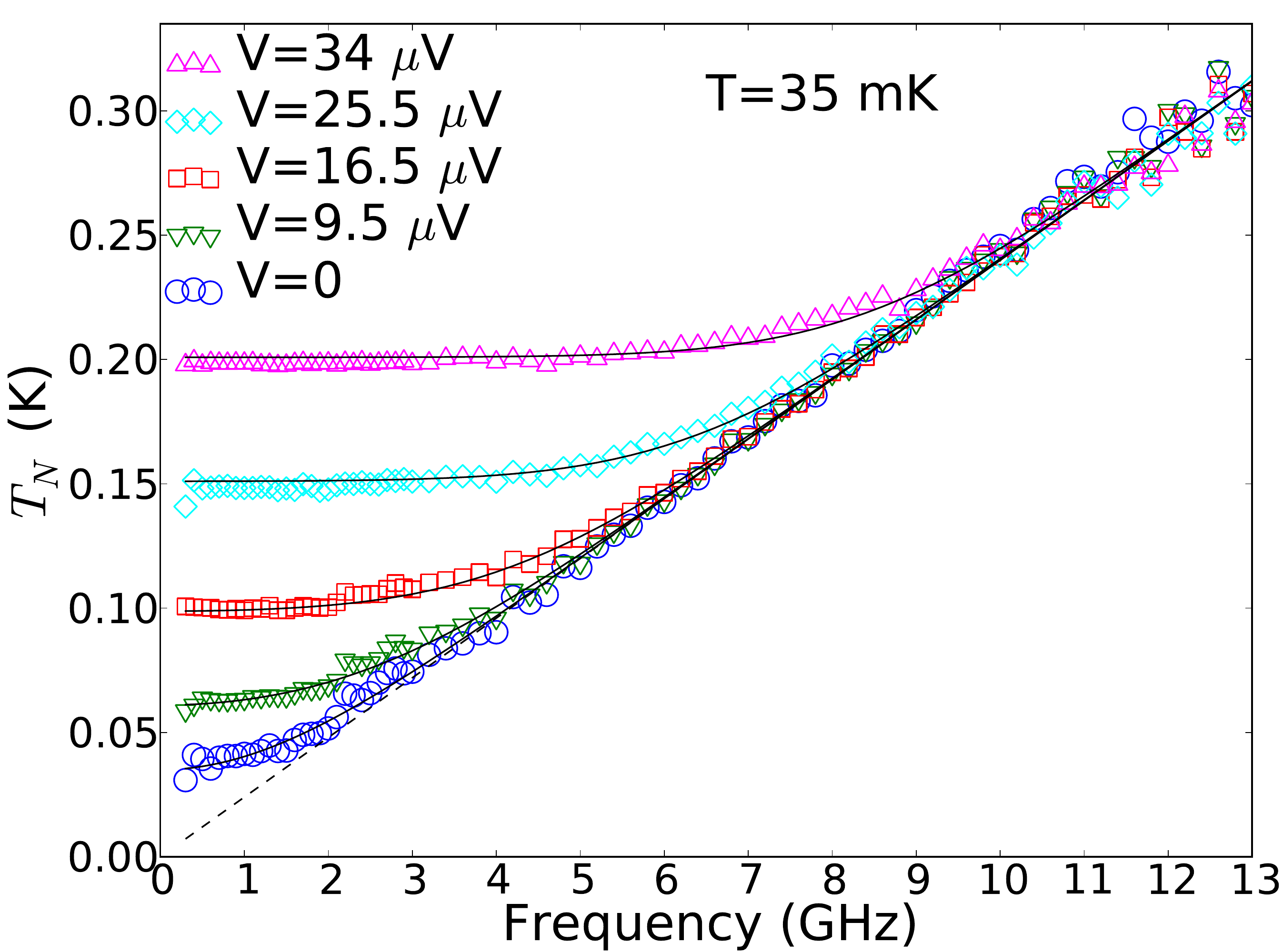}
\caption[]{\textbf{Out of equilibrium noise temperature vs. frequency for different dc voltage biases $V$ at $T = 35$ mK.} Symbols are experimental data and solid lines are theoretical expectations of equation (\ref{eq:S_V}).}
\label{fig:S_V}
\end{figure}

\emph{Shot noise spectroscopy.}
Fig. \ref{fig:S_V} shows the measurements of $T_N$ vs. frequency for various bias voltages V. The data are taken at the lowest electron temperature $T = 35$ mK. At low frequencies, \textit{i.e.} $hf < eV$, one observes a plateau corresponding to classical shot noise $S=eI$. When $hf \gg eV$, the vacuum fluctuations take over and $S=S_{vac}(f)$. Black lines on Fig. \ref{fig:S_V} are the theoretical predictions of the out of equilibrium noise spectral density \cite{Dahm1969}
\begin{align}
S(f, V, T) = \frac{1}{2} \bigg[ & S_{eq}\left( f + \frac{eV}{h}, T \right) \nonumber \\
& + S_{eq}\left( f - \frac{eV}{h}, T \right) \bigg]. 
\label{eq:S_V}
\end{align}
The data are in very good agreement with equation (\ref{eq:S_V}). Previous experiments had shown that shot noise in diffusive mesoscopic wires \cite{Schoelkopf1997} and tunnel junctions \cite{Gabelli2009} was frequency dependant by measuring the differential noise $\frac{\partial S}{\partial V}$ at specific frequencies. Here, the full spectroscopy of the absolute spectral density is obtained, which is essential to deduce the current-current correlator in time domain.

\emph{Current-current correlator in time domain.}
From the spectral density of noise measured within a very large bandwidth, it is possible to deduce the current-current correlator in the time domain by Fourier Transform (FT). The voltage dependence of the noise spectral density, given by equation (\ref{eq:S_V}), leads to a very simple form for the current-current correlator in time domain :
\begin{equation}
\label{eq:C_V}
C(t,T,V)  = C_{eq}(t,T) \cos\left( \frac{eVt}{\hbar} \right).
\end{equation}
However, since $S_{eq}(f)$ diverges as $\vert f \vert \rightarrow \infty$, its FT is not well defined so that $C_{eq}(t,T)$ diverges at all times. To circumvent this problem, we define the \textit{thermal excess noise} :
\begin{align}
\Delta S(f,T,V) = S(f,T,V) - S(f,T=0,V)
\label{eq:Delta_S}
\end{align}
which goes to zero at high frequency and is thus well suited for FT. The corresponding current-current correlator should obey :
\begin{align}
\label{eq:delta_C_V}
\Delta C(t,T,V) &= [C_{eq}(t,T) - C(t,0)]\cos\left( \frac{eVt}{\hbar} \right)\nonumber\\
&=\Delta C_{eq}(t,T) \cos\left( \frac{eVt}{\hbar} \right),
\end{align}
\begin{figure}[t]\centering
\includegraphics[width=1\columnwidth]{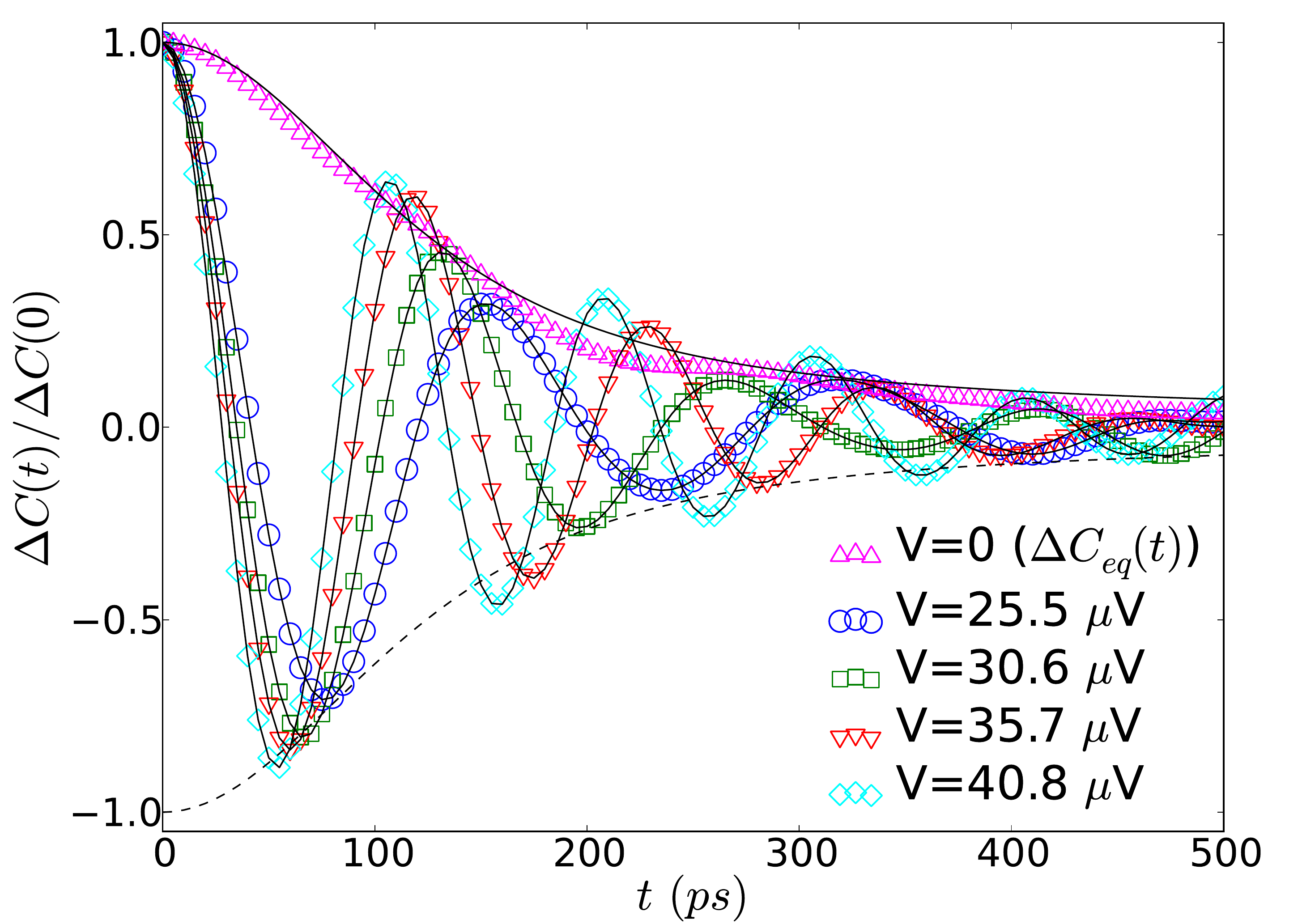}
\caption[]{\textbf{Rescaled out of equilibrium current-current correlator in time domain for five different voltages at $T= 35$ mK.} The data at V=0 correspond to the correlator at equilibrium $\Delta C_{eq}(t,T)$. Its characteristic thermal decay time is given by $\hbar/k_BT \sim 100$ ps. Solid lines are theoretical expectations.}
\label{fig:pico_C(t)}
\end{figure}
where $C(t,0)=FT[S_{vac}(f)]$ corresponds to the (infinite) jitter associated with zero point fluctuations. Note that in order to obtain such a simple and remarkable result, it is essential to subtract from $S(f,T,V)$ the noise spectral density at zero temperature but finite voltage, not $S_{vac}(f)$. The experimental equilibrium excess current-current correlator $\Delta C_{eq}(t,T)=\Delta C(t,T,V=0)$ for $T = 35$ mK was extracted by FT from the rescaled noise spectral density shown on the inset of Fig. \ref{fig:S_T} using data collected at every temperature and the scaling law we have shown. In order to avoid artificial oscillations in the data due to FT within a finite frequency range, we have used a window at frequencies between 0.3 and 12 GHz. The result is plotted on Fig. \ref{fig:pico_C(t)} ($V=0$, magenta symbols). Theoretical $\Delta C_{eq}(t,T)$ is plotted as a black line. We observe the thermal current-current fluctuations to decay with a time constant given by $\hbar/k_BT$ of $\sim$100 ps for $T=35$ mK.

\begin{figure}[t]\centering
\includegraphics[width=1\columnwidth]{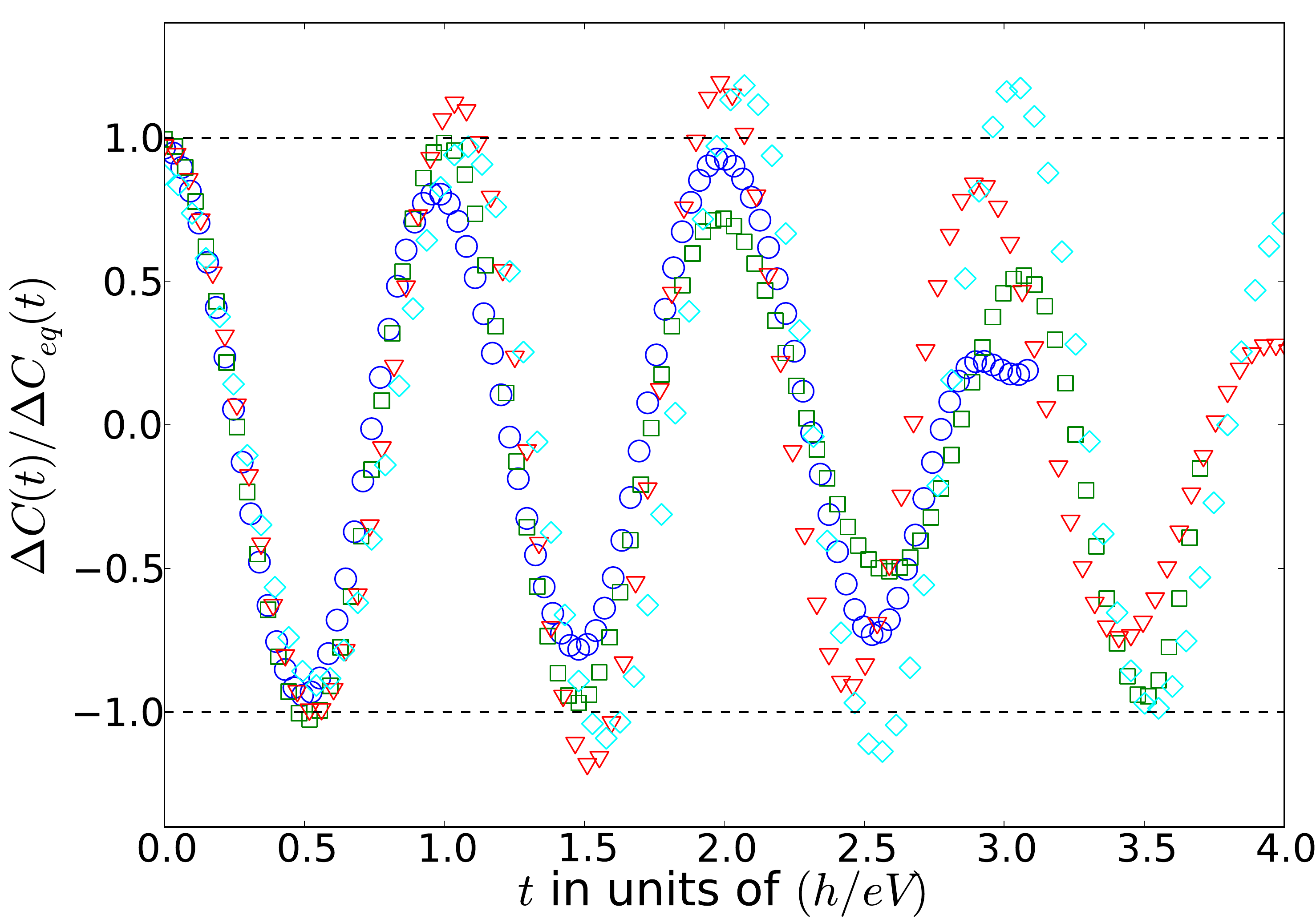}
\caption[]{\textbf{Rescaled current-current correlator in time domain vs. reduced time $eVt/h$ for various bias voltages $V=$ 25.5, 30.6, 35.7 and 40.8 $\mu V$} (same symbols as Fig. \ref{fig:pico_C(t)}).}
\label{fig:cos}
\end{figure}

Experimental data for the non-equilibrium correlator $\Delta C(t,T,V)$ at $T = 35$ mK are also shown on Fig. \ref{fig:pico_C(t)} for various voltages. One clearly observes that $\Delta C(t,T,V)$ oscillates within an envelope given by $\Delta C_{eq}(t,T)$. The period of the oscillation depends on the bias voltage. We show on Fig. \ref{fig:cos} experimental data for $\Delta C(t,V) / \Delta C_{eq}(t,T)$ as a function of the rescaled time $h/eV$. This rescaling clearly demonstrates the oscillation period being $h/eV$, in agreement with equation (\ref{eq:delta_C_V}).

\vspace{\baselineskip}
\emph{Interpretation.}
These oscillations are the result of both the Pauli principle and Heisenberg incertitude relation. To see this, let us consider a single channel conductor crossed at $t=0$ by two electrons of energy $E$ and $E'$. According to Pauli principle, the energies must be different, $E\neq E'$. But how close can $E$ and $E'$ be? According to Heisenberg incertitude relation, it takes a time $t_H \simeq h/(\vert E-E' \vert )$ to resolve the two energies, so $E$ and $E'$ cannot be considered different for times shorter than $t_H$. This means that if one electron crosses at time $t=0$, the second one must wait. Since $|E-E'|<eV$, one has $t_H > h/eV$: there is a minimum time lag $h/eV$ between successive electrons.  The regular oscillations we observe on $\Delta C$ are a direct consequence of this blockade and reflect the fact that electrons cross the sample regularly  at a pace of one electron per channel per spin direction every $h/eV$. The decay of $\Delta C(\tau)$ we observe at long time reflects the existence of a jitter which is of pure thermal origin.

Recently, the waiting time distribution (WTD) for electron transport, $W(\tau)$, has been calculated \cite{Albert2012}. For a tunnel junction, $W(\tau)$ is predicted to exhibit small oscillations with a period $\bar{\tau}= h/eV$ superposed on an eponential decay. This result is certainly closely related to our observations. It is however noteworthy that the oscillations we have observed are much more pronounced than those predicted for $W(\tau)$. 

At high bias voltage, $eV \gg k_BT, hf$, the oscillation period $h/eV$ becomes so small that the electrons no longer have to wait before tunneling. This high energy regime is the classical limit where the current flowing through the junction is characterized by a Poisson distribution. The noise spectral density is thus given by the Schottky limit $S = eI$. At low bias voltage, there are correlations between successive tunneling electrons and the resulting current distribution is no longer Poissonian. 

Our measurements were made on a tunnel junction, a device in which all conduction channels have low transmission. In the general case, equation (\ref{eq:delta_C_V}) is replaced by :
\begin{align}
\Delta C(t,T,V) =& F \,\Delta C_{eq}(t,T)\cos \left( \frac{eVt}{\hbar} \right) \nonumber \\ &+ (1-F) \Delta C_{eq}(t,T),
\end{align}
where $F$ is the Fano factor. In the case of a perfect conductor, $F=0$ and there is no oscillation of the current-current correlator, since there is no shot noise \cite{Reznikov1995,Kumar1996}. For a tunnel junction $F=1$, which corresponds to the maximal oscillations of $\Delta C(t,T,V)$.

\vspace{\baselineskip}	
\emph{Methods.}
Since we want to make an absolute measurement of the spectral density of the noise generated by the sample $S(f)$, it is necessary to calibrate both $G_A(f)$ and $S_A(f)$ at each frequency. This is achieved by making one assumption : that at high voltage $S(f)$ is given by the classical shot noise limit $S(eV \gg hf , k_BT) = eI$ \cite{Schottky1918}. Therefore, for every measurement of $S(f, V, T)$ at any frequency, voltage or temperature, we measure $P$ vs. $V$ at large voltage. From these data, we deduce the values of $G_A(f)$ and $S_A(f)$. This calibration is repeated frequently during the measurements to cancel out the drift in $G_A$ and $S_A$.

\vspace{\baselineskip}
We acknowledge fruitful discussions with W. Belzig and M. Aprili. We thank G. Laliberté for technical help. This work was supported by ANR-11-JS04-006-01, the Canada Excellence Research Chairs program, the NSERC, the MDEIE, the FRQNT via the INTRIQ, the Université de Sherbrooke via the EPIQ,  and the Canada Foundation for Innovation.

\bibliographystyle{nature}
\bibliography{CCC}

\end{document}